\documentclass{emulateapj}

\usepackage{ltgtsim}
\usepackage{epsfig}

\begin{document}

\title{Constraints on the Cardassian Expansion from the Cosmic Lens All-Sky 
Survey Gravitational Lens Statistics}

\author{J. S. Alcaniz}
\affil{Observat\'orio Nacional, Rua General Jos\'e Cristino 77, Rio de Janeiro
- RJ, 20921-400, Brazil}
\email{alcaniz@on.br}

\author{Abha Dev} 
\affil{Department of Physics and Astrophysics, University of Delhi, Delhi 
- 110007, India}
\email{abha@ducos.ernet.in}

\author{Deepak Jain} 
\affil{Deen Dayal Upadhyaya College, University of Delhi, Delhi - 110015, 
India}
\email{deepak@physics.du.ac.in}

\begin{abstract}
The existence of a dark energy component has usually been invoked
as the most plausible way to explain the recent observational
results. However, it is also well known that effects arising from
new physics (e.g., extra dimensions) can mimic the gravitational
effects of a dark energy through a modification of the Friedmann
equation. In this paper we investigate some observational
consequences of a flat, matter dominated and
accelerating/decelerating scenario in which this modification is
given by $H^{2} = g(\rho_m, n, q)$ where $g(\rho_m, n, q)$ is a
new function of the energy density $\rho_m$, the so-called
generalized Cardassian models. We mainly focus our attention on
the constraints from the recent Cosmic All Sky Survey (CLASS)
lensing sample on the parameters $n$ and $q$ that fully
characterize the models. We show that, for a large interval of the
$q - n$ parametric space, these models are in agreement with the
current gravitational lenses  data. The influence of these
parameters on the acceleration redshift, i.e., the redshift at
which the universe begins to accelerate, and on the age of the universe 
at high-redshift is also discussed.
\end{abstract}

\keywords{Cosmology: theory --- dark matter --- distance scale}

\section{Introduction}

Over the last years, a considerable number of high quality observational data 
have transformed radically the field of cosmology. Results from distance 
measurements of type Ia supernovae (SNe Ia) (Perlmutter et al. 1999; Riess et 
al. 1998; 2004) combined with Cosmic Microwave Background (CMB) observations 
from ballon (de Bernardis et al. 2000;  O'Dwyer et al. 2003), ground 
(Mason et al. 2003) and satelite experiments (Spergel et al. 2003) and with 
dynamical estimates of the quantity of matter in the universe (Calberg et al. 
1996; Dekel et al. 1997) seem to indicate that the simple picture provided by 
the standard cold dark matter scenario (SCDM) is not enough. These observations 
are usually explained by introducing a new hypothetical energy component with 
negative pressure, the so-called dark energy or {\emph{quintessence}} 
(For recent reviews on this topic, see Sahni \& Starobinsky, 2000; Peebles \& Ratra 2003; Padmanabhan, 
2003; Lima, 2004 and references therein). Besides its consequences on
fundamental physics, if confirmed, the existence of this 
dark component would also provide a definitive piece of information connecting 
the inflationary flatness prediction with astronomical data. 

On the other hand, 
it is also well known that not less exotic mechanisms like, e.g., geometrical 
effects from extra dimensions may be capable of explaning such observational 
results. The basic idea behind these ``braneworld cosmologies" is that our 
4-dimensional Universe would be a surface or a brane embedded into a higher 
dimensional bulk space-time to which gravity could spread (Ho\v{r}ava \&
Witten, 
 1996a; 1996b; Randall L. \& Sundrum, 1999; Dvali, Gabadadze \& Porrati, 
2000; Perivolaropoulos \& Sourdis 2002; Maia {\it et al.}, 2004). In some of these scenarios the observed 
acceleration of the Universe can be explained (without dark energy) from the
fact 
that the bulk gravity sees its own curvature term on the brane acting as a 
negative-pressure dark component which accelerates the Universe. A 
natural conclusion from these and other similar studies is that dark energy, or 
rather, the gravitational effects of a dark energy could actually be achieved 
from a modification of the Friedmann equation arising from new physics.

Following 
this reasoning, several authors have recently investigated cosmologies with a 
modified Friedmann equation from extra dimensions as an alternative explanation 
for the recent observational data (Deffayet, Dvali \& Gabadadze, 2002;
Alcaniz, 2002; Jain, 
Dev \& Alcaniz, 2002; Alcaniz, Jain \& Dev, 2002; Lue \& Starkman,
2003; Lue, Scoccimarro \& 
Starkman, 2004; Alcaniz \& Pires, 2004; Zhu \& Alcaniz, 2004; Sahni \& 
Shtanov 2004). For example, in
Sahni \& 
Shtanov (2002) a new class of braneworld models which admit a 
wider range of possibilities for dark energy than do the usual quintessence 
scenarios was investigated. For a subclass of the parameter values, the 
acceleration of the universe in this class of models can be a transient
phenomena 
which could help reconcile an accelerating universe with the requirements 
of string/M-theory (Fischler et al. 2001).

Recently, Dvali \& Turner (2003)
explored 
the phenomenology and detectability of a correction on the Friedmann equation
of 
the form $(1 -\Omega_m)H^{\alpha}/H_o^{\alpha - 2}$, where $\Omega_m$ is 
the matter density parameter, $H$ is the Hubble parameter (the subscript ``o" 
refers to present time) and $\alpha$ is a parameter to be adjusted by the 
observational data. Such a correction behaves like a dark energy with an 
effective equation of state given by $\omega = -1 + \alpha/2$ in the recent
past 
and like a cosmological constant ($\omega = -1$) in the distant future. Also 
based on extra dimensions physics, Freese \& Lewis (2002) proposed the
so-called 
{\emph{Cardassian expansion}}, a model in which the universe is flat, matter 
dominated and currently accelerated. In the Cardassian universe, the new 
Friedmann equation is given by $H^{2} = g(\rho_m)$, where $g(\rho_m)$ is 
an arbitrary function of the matter energy density $\rho_m$. In the first
version 
of these scenarios, the function $g(\rho_m)$ was given by  $g(\rho_m) = A\rho_m
+ 
B\rho_m^{n}$, with the second term driving the acceleration of the universe at
a 
late epoch after it becomes dominant (a detailed discussion for the origin of
this 
Cardassian term from extra dimensions physics is given by Chung \& Freese 2002.
See also Cline \& Vinet 2003). 
Although being completely different from the physical viewpoint, it was
promptly 
realized that by identifying some free parameters Cardassian models and 
quintessence scenarios parameterized by an equation of state $p = \omega \rho$ 
predict the same observational effects in what concerns tests involving only
the 
evolution of the Hubble parameter with redshift (Freese \& Lewis 2002).

More recently, several forms for the function $g(\rho_m)$ have been proposed 
(Freese 2003). In particular, Wang {\it et al.} (2003) have studied some 
observational characteristics of a direct generalization of the 
original Cardassian model. According to the authors, the 
observational expressions in this new scenario are very different from 
generic quintessence cosmologies and fully determined by two 
dimensionless parameters, $n$ and $q$. Observational constraints from a variety
of 
astronomical data have also been investigated recently, both in the original 
Cardassian model (Zhu \& Fujimoto 2002; 2003; Sen \& Sen 2003a; 2003b) and in
its generalized 
versions (Wang et al. 2003; Savage, Sugiyama \& Freese 2004; Amarzguioui,
Elgaroy \& Multamaki, 2004). 
Perhaps the most interesting feature of Cardassian models is that although
being matter 
dominated, they may be accelerating and may still reconcile the indications for
a 
flat universe ($\Omega_{\rm{total}} = 1$) from CMB observations with
clustering estimates that point consistently to $\Omega_m \simeq
0.3$ with no need to invoke either a new dark component or a
curvature term. In these scenarios, it happens through a
redefinition of the value of the critical density (Freese \& Lewis 2002) -- see
also below.

The aim of this paper is to test the viability of generalized Cardassian (GC)
scenarios from statistical properties of gravitational lensing. To this end,
we use the recent Cosmic All Sky Survey (CLASS) lensing sample, which 
constitutes the so far largest lensing sample suitable for statistical analysis. 
We show that for a large interval of the parameters $n$ and $q$ characterizing 
the models, GC scenarios are fully compatible with the current gravitational 
lensing observations. We also explore other observational aspects of the model 
like, e.g., those related to the acceleration redshift (i.e., the redshift at 
which the universe begins to accelerate) and to the age-redshift relation.

\section{The generalized cardassian scenario}

The modified Friedmann equation for GC models is (Wang et al. 2003)
\begin{eqnarray}
H^{2} =  \frac{8\pi G \rho_m}{3}\left[1 + 
\left(\frac{\rho_{card}}{\rho_m}\right)^{q(1 - n)}\right]^{1/q}, 
\end{eqnarray} 
where $n$ and $q$ are two free parameters to be adjusted by the observational 
data, $\rho_{{card}} =\rho_o(1 + z_{{card}})^{3}$ is the energy density at
which 
the two terms inside the bracket become equal and $\rho_o$
is the present day matter density. Note that for values of $n = 0$ and $q = 1$, 
GC scenarios reduce to Cold Dark Matter models with a cosmological constant 
($\Lambda$CDM). As explained in Wang et al. (2003), the first term inside the
bracket 
dominates initially in such a way that the standard picture of the early 
universe is completely maintained. At a redshift $z_{{card}}$, these two terms 
become equal and, afterwards, the second term dominates, leading or not to an 
accelerated expansion (note, however, that $z_{card}$ is not necessarily 
the acceleration redshift. See, e.g., the discussion below on the deceleration
parameter).

Evaluating Eq. (1) for present day quantities we find
\begin{equation}
H_o^{2} = \frac{8\pi G \rho_o}{3}\left[1 + (1 + z_{card})^{3q(1 - 
n)}\right]^{1/q}. 
\end{equation} 
From this equation, we see that $\rho_o$ is also the critical density that now 
can be written as 
\begin{equation}
\rho_o = \rho_{c,old} \times \left[1 + (1 + z_{card})^{3q(1 - 
n)}\right]^{-{1/q}}, 
\end{equation} 
where $\rho_{c,old} = 1.88 \times 10^{-29}h^{2}\rm{gm/cm^{3}}$ is the standard 
critical density and $h$ is the present day Hubble parameter in units of 100 
${\rm{km.s^{-1}Mpc^{-1}}}$. Note that for some combinations of the parameters 
$z_{card}$, $n$ and $q$ the critical density can be much lower than the 
one previously estimated. In other words it means that in the context of GC 
models it is possible to make the dynamical estimates of the quantity of matter 
that consistently point to $\rho_o \simeq (0.2-0.4)\rho_{c,old}$ compatible
with 
the observational evidence for a flat universe from CMB observations and the 
inflationary flatness prediction with no need of a dark energy component (see 
Freese \& Lewis (2002) for a more detailed discussion). In Fig. 1 we show a
generalized 
version of the Figure 1 of Freese \& Lewis (2002) in which the plane $z_{card}
- n$ is 
displayed for selected values of $q$. The contours are labeled indicating the 
fraction of the standard critical density for different combinations of 
$z_{card}$ and $n$. In particular, the points inside the shadowed area
delimited 
by the contours 0.2 - 0.4 are roughly consistent with the present clustering 
estimates (Calberg et al. 1996; Dekel et al. 1997). 

\begin{figure}
\centerline{\psfig{figure=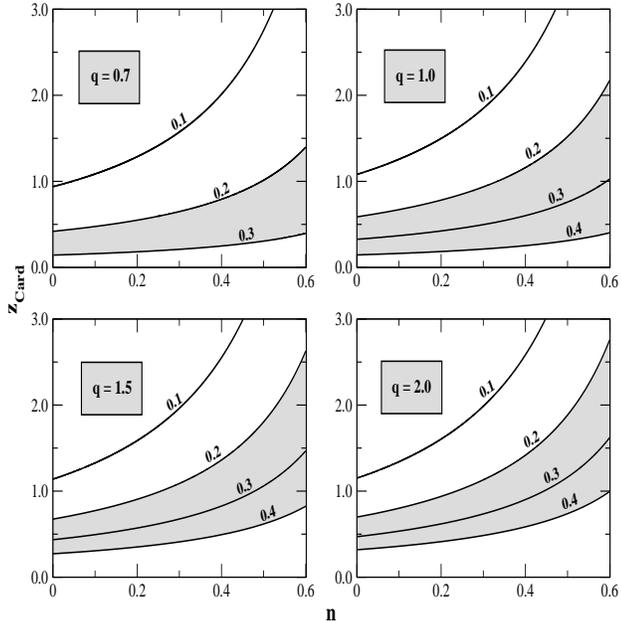,width=3.5truein,height=3.3truein,angle=-90}
\hskip 0.1in}
\caption{$z_{card} - n$ diagrams for the ratio $\rho_o/\rho_{c,old}$ (Eq. 4) 
and selected values of $q$. The contours are labeled indicating the 
corresponding fraction of the standard critical density.  Points inside the 
shadowed area are roughly consistent with the present clustering estimates 
(Calberg {\it et al.}, 1996).} 
\end{figure}

From Eq. (3), we see that the observed matter density parameter in GC models
can be written as 
\begin{equation} 
\Omega_m^{obs} = 
\frac{\rho_o}{\rho_{c,old}} = \left[1 + (1 + z_{card})^{3q(1 - 
n)}\right]^{-{1/q}}. 
\end{equation} 
For the GC expansion parameterized by $n$ and $q$, the deceleration parameter
as a function of the 
redshift has the following form
\begin{equation}
q(z) \equiv - \frac{\ddot{R}R}{\dot{R}^{2}} = -1 + 
\frac{1}{2}\frac{d{\rm{ln}}f^{2}(z, \Omega_m^{obs}, q,n)}{d{\rm{ln}}(1 + z)},
\end{equation}
where an overdot denotes derivative with respect to time, $R(t)$ is the
cosmological scale 
factor and $f(z,\Omega_m^{obs}, q, n)$ is given by Eq. (7). The behavior of the 
deceleration parameter as a function of redshift for some selected values of
$n$ and $q$ and $\Omega_m^{obs} = 0.3$ is shown in Figure 2. 
As discussed earlier, although completely dominated by matter, GC scenarios
allow periods of accelerated 
expansion for some combinations of the parameters $n$ and $q$. Note that the 
present acceleration is basically determined by the value of $n$ and that the 
smaller its value the more accelerated is the present expansion for a given 
value of $q$. For example, for $q = 1.5$ and $n = 0$, GC models accelerate 
presently faster ($q_o \simeq -0.75$) than flat $\Lambda$CDM scenarios with 
$\Omega_{\Lambda} =0.7$ ($q_o\simeq -0.5$) although the acceleration redshift
is 
almost identical ($z_a \simeq 0.7$) while for the same value of $q$ and $n =
0.5$ 
we find $q_o \simeq -0.13$ and $z_a \simeq 0.52$. In order to make clear the 
difference between $z_{card}$ and $z_a$, for the latter values of $q$ and $n$, 
we find directly from Eq.(4) $z_{card} = 1.06$. It means that although becoming 
dominant at $z_{card} \simeq 1$ the second term of Eq. (1) will drive 
an accelerated expansion only $\sim 1.7$ Gyr later, at $z_a \simeq 0.52$.

\begin{figure}
\centerline{\psfig{figure=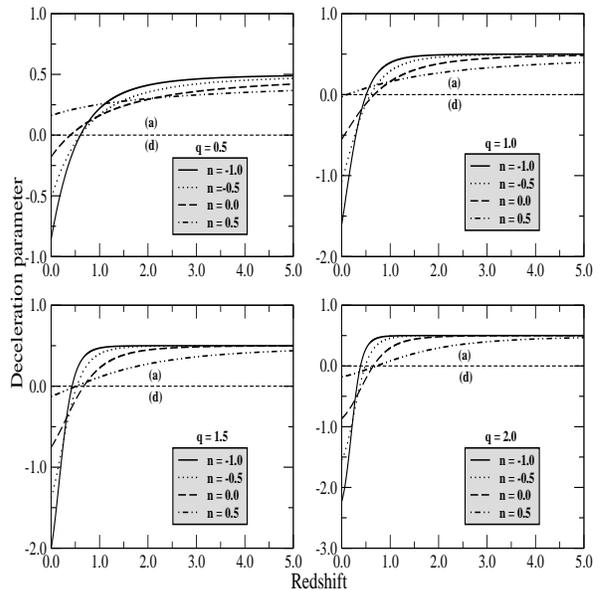,width=3.5truein,height=3.3truein,angle=-90}
\hskip 0.1in} 
\caption{Deceleration parameter as a function of redshift for some selected 
values of $n$ and $q$ and $\Omega_m^{obs} = 0.3$. The horizontal line labeled 
(a)/(d) ($q_o= 0$) divides models with a accelerating (a) or decelerating (d)
expansion at a given redshift. As discussed in the text, the present value of 
$q_o$ is basically determined by the value of $n$.} 
\end{figure}

From the above equations, it also is straightforward to show that the 
age-redshift relation is now given by (the total expanding age of the Universe 
$t_o$ is obtained by taking $z = 0$) 
\begin{equation}
t_z = \frac{1}{H_o}\int_{o}^{x'} {dx \over x f(x, \Omega_m^{obs}, q, n)},
\end{equation}
where  $x' = {R(t) \over R_o} = (1 + z)^{-1}$ is a convenient integration 
variable and the dimensionless function $f(x, \Omega_m^{obs}, q, n)$, obtained 
from Eqs. (1) and (4), is written as 
\begin{eqnarray}
f(x, \Omega_m^{obs}, q, n)  =  \left\{\frac{\Omega_m^{obs}}{x^{3}}\left[1 + 
\frac{(\Omega_m^{obs})^{-q} -1}{x^{-3q(1
- n)}}\right]^{1/q}\right\}^{\frac{1}{2}}.
\end{eqnarray}
An interesting study on the total expanding age of the Universe in generalized
Cardassian scenarios was 
recently presented by Savage et al. (2004). In light of first year WMAP data,
it was shown that for
a subset of Cardassian models the age of the universe varies in
the interval 13.4 - 13.8 Gyr, which is surprisingly close to the
standard $\Lambda$CDM prediction ($13.7 \pm 0.2$ Gyr).

The angular diameter distance to a light source, defined as the ratio of the 
source diameter to its angular diameter, is an important concept in lensing 
statistics. For the class of GC models here investigated, the angular 
diameter distance, $D_{LS}(z_L, z_S) = {R_or_1(z_L,z_S)/(1 + z_S)}$, between two 
objects, for example a lens at $z_L$ and a source (galaxy) at $z_S$, 
reads 
\begin{eqnarray} 
D_{LS}(z_L, z_S) & = & \frac{H_o^{-1}}{(1 + z_S)} 
\int_{x'_S}^{x'_L} {dx \over x^{2} f(x, \Omega_m^{obs}, q, n)} . 
\end{eqnarray}

\section{Lensing constraints using CLASS Sample}

In this Section we present our study of statitical properties of gravitational
lenses in GC scenarios based 
on the final CLASS well-defined statistical sample (for similar studies 
in quintessence and quartessence scenarios see, e.g., Chae et al. 2004; Jain et al. 2004). 

The CLASS sample consists of 8958 radio sources out of which 13 sources are multiply 
imaged. Here we work only with those multiply imaged sources whose 
image-splittings are known (or likely) to be caused by single galaxies. There 
are 9 such radio sources: 0218+357, 0445+123, 0631+519, 0712+472, 
0850+054, 1152+199, 1422+231, 1933+503, 2319+051. We, therefore, following Chae et 
al. (2002) and Dev, Jain \& Mahajan (2004), work with a total of 8954 radio 
source. The sources probed by CLASS at $\nu = 5$ GHz are well represented by 
power-law differential number-flux density relation: $\left |dN/dS \right| 
\propto (S/S^{0})^{\eta}$ with $\eta = 2.07 \pm 0.02$ ($1.97 \pm
0.14$) for $S \geq S^{0}$ ($ \leq S^{0}$) where $S^{0} = 30$ mJy (Browne et
al., 2003). The CLASS unlensed sources can be adequately described by a
Gaussian model with  mean 
redshift $z = 1.27$ and a dispersion of $0.95$.

In our analysis we assume the singular isothermal sphere
(SIS) model for the lens mass distribution  (Turner, Ostriker \& Gott 1984).
For
the present analysis we also ignore the evolution of the number
density of galaxies and assume that the comoving number density is
conserved and currently given by 
\begin{equation} \label{number}
n_o = \int_0^{\infty} \phi(L) dL,
\end{equation}
where  $\phi (L)$ is the well known Schechter Luminosity Function (LF). The
differential optical depth of 
lensing in traversing $dz_L$ with angular separation between $\phi$ and $\phi +
d\phi$ is (Fukugita, 
Futamase \& Kasai, 1990, Turner, 1990; Fukugita et al., 1992)
\begin{eqnarray}
\frac{d^{2}\tau}{dz_{L}d\phi}d\phi dz_{L} &=& F^{*}\,(1 +
z_{L})^{3}\,\left({{D_{OL} D_{LS}}\over{ R_{o}
D_{OS}}}\right)^{2}\ \frac{1}{R_{o}}\, \frac{cdt}{dz_{L}}
\times \nonumber \\ && \times
\frac{\gamma/2}{\Gamma(\alpha+1+\frac{4}{\gamma})}
\left(\frac{D_{OS}}{D_{LS}}\phi\right)^{\frac{\gamma}{2}(\alpha+1+\frac{4}{
\gamma})} 
\nonumber \\ && \times
{\rm{exp}}\left[-\left(\frac{D_{OS}}{D_{LS}}\phi\right)^{\frac{\gamma}{2}
}\right]\frac{d\phi}{\phi} dz_{L}, 
\label{diff}
\end{eqnarray}
where the function $F^{*}$ is defined as
\begin{equation}
F^* = {16\pi^{3}\over{c\, H_{0}^{3}}}\phi_\ast
v_\ast^{4}\Gamma\left(\alpha +{4\over\gamma} +1\right).
\end{equation}

\noindent In Eq. (10) $D_{OL}$, $D_{OS}$ and $D_{LS}$ are, respectively,
the angular diameter distances from the observer to the lens, from
the observer to the source and between the lens and the source (see Eq. 8). In
order to relate the characteristic luminosity $L_*$ to the
characteristic velocity dispersion $v_{*}$, we use the
Faber-Jackson relation (Faber \& Jackson, 1976) for early-type galaxies ($L_*
\propto {v_*}^{\gamma}$), with $\gamma = 4$. For the analysis
presented here we neglect the contribution of spirals as lenses
because their velocity dispersion is small when compared to ellipticals.

Two large-scale galaxy surveys, namely, the 2dF Galaxy 
Redshift-Survey (2dFGRS)\footnote{http://msowww.anu.edu.au/2dFGRS/} and the 
Sloan Digital Sky Survey (SDSS)\footnote{http://www.sdss.org/} have produced 
converging results on the total LF. These surveys determined the 
Schechter parameters for galaxies (all types) at $z \le 0.2$. For our analysis
here, we adopt the 
normalization corrected Schechter parameters of the 2dFGRS survey 
(Folkes {\it et al.} 1999; Chae 2002): $\alpha = -0.74$, $\phi^{*} = 0.82
\times 10^{-2} 
h^3\mathrm{Mpc^{-3}}$, $v^{*} = 185\,\mathrm{km/s}$ and $F^{*} = 0.014$.

The total optical depth $\tau(z)$ along the line of sight from an
observer at $z = 0$ to a source at $z_S$ is given by
\begin{equation}
\tau(z_S) = \frac{F^{*}}{30} \left[D_{OS}(1+z_{S})\right]^{3}
R_{o}^{3}. \label{atau}
\end{equation}
where $D_{OS}$ is given by Eq. (8). Figure 3 shows, for the fixed
value of $\Omega_m^{obs} = 0.3$,  the normalized optical depth as
a function of the source redshift for values of $q = $ 0.3, 1.0,
2.0 and 3.0 and $n = $ -1.0, -0.5, 0.0 and 0.5. Note that a
decrease in the value of $n$ at fixed $\Omega_m^{obs}$ and $q$
tends to increase the optical depth for lensing. For $q = 1.0$ at
$z_S = 3.0$, the value of $\tau/F^{*}$ for $n = 0.5$ is down from
that one for $n = -1.0$ by a factor of $\sim 2.02$, while the same
values of $n$ and $q = 2.0$ provide values for $\tau/F^{*}$ that
are up from the previous ones only by a factor of $\sim$ 1.1 and
1.25, respectively. It clearly shows that the optical depth is a
more sensitive function to the parameter $n$ than to the index
$q$. As commented earlier, this particular feature is also noted
for the other analyses discussed in this paper.

\begin{figure}
\centerline{\psfig{figure=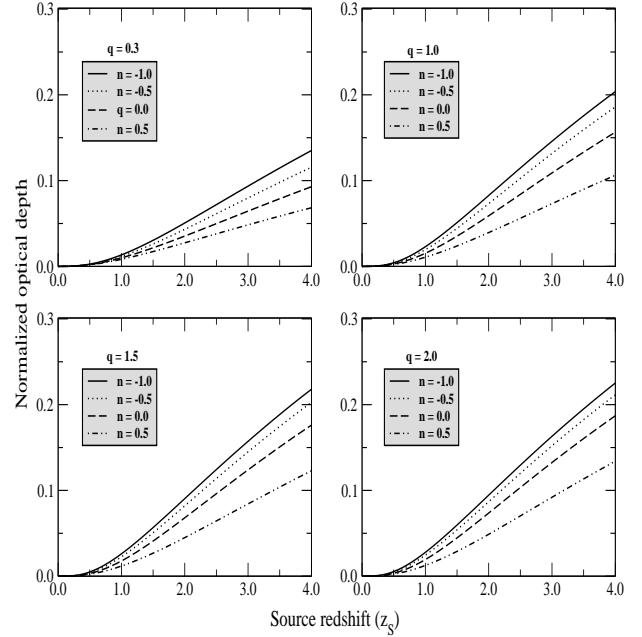,width=3.5truein,height=3.3truein,angle=-90}
\hskip 0.1in} 
\caption{Normalized optical depth ($\tau/F^{*}$) as a function of the source 
redshift $z_S$ for some selected  valuesof $n$ and $q$ and a fixed value 
of $\Omega_m^{obs} = 0.3$.} 
\end{figure}

The normalized image angular separation distribution for  a source at $z_{S}$
is 
obtained by integrating $\frac{d^{2}\tau}{dz_{L}\,d\phi}$ over $z_{L}$, i.e., 
\begin{equation}
{d{\mathcal{P}}\over d\phi}\, =\,
{1\over\tau(z_S)}\int_{0}^{z_{s}}\,{\frac{d^{2} \tau
}{dz_{L}d\phi}} {dz_{L}}. 
\end{equation}
The corrected (for magnification and selection effects) image separation
distribution function for a single 
source at redshift $z_{S}$ is given by (Kochanek, 1996; Chiba \& Yoshii, 2001)
\begin{eqnarray} 
P'(\Delta\theta)\,& =& \, \mathcal{B}\,{{{\gamma}} \over 2\, \Delta \theta}
\int_{0}^{z_S}\left 
[{D_{OS}\over{D_{LS}}} \phi \right ]^{{\frac{\gamma}{2}}(\alpha + 1+ 
{\frac{4}{\gamma}})} \,F^{*}\,{cdt\over dz_{L}} \times \nonumber \\ && \times 
\exp\left[- \,\left({D_{OS}\over{D_{LS}}}\phi\right)^{\frac{\gamma}{2}}\right] 
{(1 + z_{L})^{3} \over\Gamma\left(\alpha +{4\over\gamma} +1\right)}  
\nonumber \\ &&\times 
\left[\,\left ({D_{OL}D_{LS}\over R_0
  D_{OS}}\right)^{2}\,{1\over R_0}\right ]\,\,{dz_{L}}.
\label{dist}
\end{eqnarray}

\noindent Similarly, the corrected lensing probability for a given
source at redshift $z$ can be written as
\begin{equation}
P' = \tau(z_S) \,\int {d{\mathcal{P}}\over d\phi} \mathcal{B}\; d\phi, 
\label{prob}
\end{equation}
where $\phi$ and $\Delta\theta$ are related to as $\phi =
{\frac{\Delta\theta}{8 \pi (v^{*}/c)^2}}$ and $\mathcal{B}$ is the
magnification bias. It is worth emphasizing that this is considered because, as 
widely known, gravitational lensing causes a magnification of images and this
transfers the  lensed sources to higher flux density bins. In
other words, the lensed sources are over-represented in a
flux-limited sample.

The magnification bias  ${\mathcal
B}(z_S,S_\nu)$ increases the lensing probability significantly in
a bin of total flux density ($S_\nu$) by a factor 
\begin{eqnarray}
\mathcal{B}(z_S,S_\nu) \,& =& \,  \left
|\frac{dN_{z_S}(>S_\nu)}{dS_\nu} \right|^{-1} \\ && \times
\int_{\mu_{min}}^{\mu_{max}} \left
|\frac{dN_{z_S}(>S_\nu/\mu)}{dS_\nu}\,p(\mu)\right |{1\over\mu}
\,d\mu.  \nonumber
\label{B2}
\end{eqnarray}
Here  $N_{z_{S}}(> S_\nu)$ is the intrinsic flux density relation for 
the source population at redshift $z_{S}$. $N_{z_{S}}(> S_\nu)$ 
gives the number of sources at redshift $z_{S}$ having 
flux greater than $S_\nu$. For the SIS model, the magnification 
probability distribution is $p(\mu) = 8/{\mu}^{3}$. The minimum ($\mu_{min}$)
and maximum ($\mu_{max}$) total magnifications in Eq. (16) depend on 
the observational characteristics as well as on the lens model. For the SIS 
model, the minimum total magnification is $\mu_{min} \simeq 2$ while $\mu_{max} 
= \infty$. The magnification bias $\mathcal B$ depends on the differential 
number-flux density relation $\left|dN_{z_{S}}(> S_{\nu})/dS_{\nu}\right|$, 
which means that such a relation needs to be known as 
a function of the source redshift. Since, at present, redshifts of only a few 
CLASS sources are known, we ignore redshift dependence of
the differential number-flux density relation. Following Chae (2002), we 
further ignore the dependence of the differential number-flux density relation 
on the spectral index of the source.

Finally, we emphasize that two important selection criteria for CLASS 
statistical sample are (i) that the ratio of the flux densities of the fainter
to the brighter images ${\mathcal R}_{min}$ is $ \ge 0.1$. Given such an
observational limit, the minimum total magnification for double
imaging for the adopted model of the lens is $\mu_{min} = 2 
(1+\mathcal{R}_{min}/1-\mathcal{R}_{min})$; (ii) that the image 
components in lens systems must have separations $\ge 0.3$ ~ arcsec. We 
incorporate this selection criterion by setting the lower limit of
$\Delta\theta$ in equation (\ref{prob}) as $0.3$ arcsec.

We perform a maximum-likelihood analysis to determine the
confidence levels in the $n - q$ space. The likelihood function is defined by 
\begin{equation}
{\cal{L}} =  \prod_{i=1}^{N_{U}}(1-P^{'}_{i})\,\prod_{k=1}^{N_{L}}
P'(\Delta\theta), 
\label{LLF}
\end{equation}
where $N_{L}$ is the observed number of multiple-imaged lensed
radio sources and $N_{U}$ is the number of unlensed sources in the
adopted sample. $P_{i}^{'}$ is the probability of the $i^{th}$
source to get lensed and $P_{k}^{'}(\Delta\theta)$ is the probability
of the $k^{th}$ source to get lensed with the observed image
separation $\Delta\theta$. 

\begin{figure}
\vspace{.2in}
\centerline{\psfig{figure=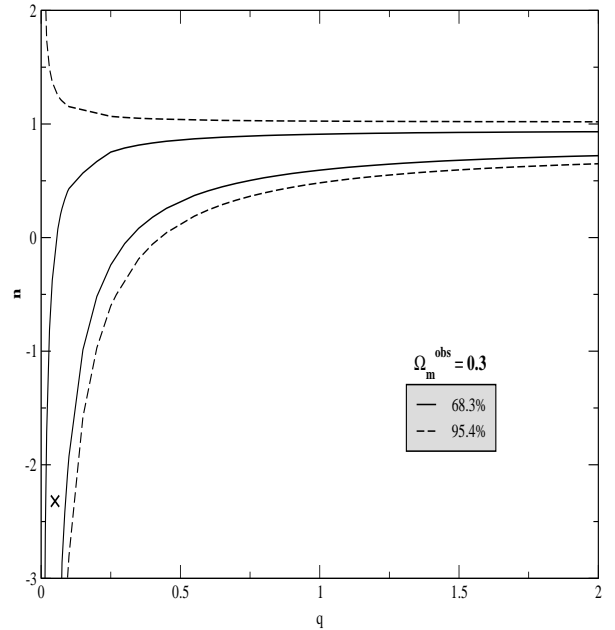,width=3.5truein,height=3.3truein,angle=-90}
\hskip 0.1in} 
\caption{Confidence regions in the plane $n - q$ arising from lensing statistics 
for a fixed value of $\Omega_m^{obs} = 0.3$. Solid (dashed) lines indicate 
contours of constant likelihood at $68.3\%$ c.l. ($95.4\%$ c.l.). The best-fit 
model is indicated by ``$\times$".} 
\end{figure}

Figure 4 shows the main results of our analysis. There, we show contours of
constant likelihood ($68.3\%$ and
$95.4\%$) in the parameter space $n - q$. From the above equation
we find that the maximum value of the likelihood function is
located at $n =-2.32$ and $q = 0.05$, which means that the cardassian term of 
Eq. (1) takes over only recently, i.e., at $z_{card} \simeq
{\cal{O}}(10^{-3})$. 
At the 1$\sigma$ level, however, the entire range of $q$ (if we
consider for instance $0 < q \leq 2$) is compatible with the
observational data for a fixed value of $\Omega_{m}^{obs} = 0.3$ (a similar 
conclusion also holds for small variations on the value of $\Omega_{m}^{obs} = 
0.3$, i.e., $\pm 0.1$, which is inside the uncertainties of current clustering
estimates). As 
observed earlier, this result suggests that a large class of GCs cenarios is in 
accordance with the current gravitational lensing data. This best-fit scenario 
also corresponds to a $7.7h^{-1}$-Gyr-old universe ($z = 0$) and provides an
universe old enough to accommodate some recent age estimates of high-$z$
objects. For example, at $z = 1.55$ and $z = 1.43$ Eq. (6) provides\footnote{For 
$H_o =64$ ${\rm{kms^{-1}Mpc^{-1}}}$, which corresponds to $1\sigma$ lower bound 
obtained by the $HST$ key project (Freedman et al. 2003).} 
$t_z \simeq 3.5$ Gyr and $t_z \simeq 3.7$ Gyr, respectively, i.e., values that 
are in agreement with the age estimates for the radio galaxies LBDS 53W091 and 
LBDS 53W069 (Dunlop et al. 1996; Dunlop 1988; Alcaniz \& Lima 1999). For the 
recent discovery of the quasar APM 08279+5255 (Hasinger \& Komossa 2002; Komossa 
\& Hasinger 2003) at $z = 3.91$, however,
these values of $q$ and $n$ provide $t_z \simeq 1.4$ Gyr while the age
estimate for this object lies between 2.0 - 3.0 Gyr. A similar
problem is also faced by several cosmological scenarios, including the current 
concordance $\Lambda$CDM model (see Fria\c{c}a, Alcaniz \& Lima, 2004).

\section{Conclusion}

The possibility of an accelerating universe from distance
measurements of type Ia supernovae constitutes one of the most
important results of modern cosmology. These observations
naturally lead to the idea of a dominant dark energy component
with negative pressure since all known types of matter with
positive pressure generate attractive forces and decelerate the
expansion of the universe [$\ddot{R} \propto - (\rho + 3p$)]. On the other
hand, the realization that
dark energy or the effects of dark energy could be a manifestation
of a modification to the Friedmann equation arising from extra
dimension physics has opened up an unprecedented opportunity to
establish a more solid connection between particle physics and
cosmology. Many \emph{braneworld scenarios} have been proposed in the
recent literature with some of them presenting interesting
features which make them a natural alternative to the standard
model. Here we have analyzed some observational consequences of
one of these scenarios, the so-called generalized cardassian
expansion, recently proposed by Wang et al. (2003). We have studied
the observational constraints on the parameters $n$ and $q$ that
fully characterize the model from statistical properties of
gravitational lensing. To this end, we have used the final CLASS well-defined
statistical sample, which constitutes the most recent gravitational lensing
data. From our analysis we have found that at
$1\sigma$ level a large portion of the parametric space $q - n$ is in agreement
with the current gravitational lensing observations, with the maximum of the
likelihood function located at $n =-2.32$ and $q = 0.05$ which corresponds to
a $7.7h^{-1}$-Gyr-old universe and $z_{card} \simeq {\cal{O}}(10^{-3})$. These
and other similar results show that there exist viable and interesting
alternatives to our current standard cosmological model ($\Lambda$CDM) based
on extra dimension physics. Further observational analysis on the generalized
cardassian scenario will appear in a forthcoming communication.

\acknowledgments The authors are very grateful to Zong-Hong Zhu and Carlos da
Silva Vilar for helpful 
discussions. JSA is supported by the Conselho Nacional de Desenvolvimento
Cient\'{\i}fico e Tecnol\'{o}gico (CNPq - Brasil) and CNPq (62.0053/01-1-PADCT 
III/Milenio).

\end{document}